\documentclass{article}
\usepackage{spconf,amsmath,graphicx,hyperref}
\usepackage{algorithm}
\usepackage{algpseudocode}
\usepackage{comment}
\usepackage{bm,amsfonts,amssymb}
\usepackage{xcolor}
\usepackage{bbold}
\usepackage{multirow}
\usepackage[nameinlink, capitalise]{cleveref}
\usepackage{arydshln}
\usepackage{booktabs}

\newcommand{\bhline}[1]{\noalign{\hrule height #1}}

\title{Chunkwise Aligners for Streaming Speech Recognition}
\name{\begin{tabular}{c} Wen Shen Teo$^{1,2 \ast}$\thanks{$^{\ast}$ Equal contribution. This work was conducted during the first author’s internship at NTT, Inc.}, Takafumi Moriya$^{2 \ast}$, Masato Mimura$^{2}$ \end{tabular}}
\address{\begin{tabular}{c} $^{1}$University of Electro-Communications, Japan; $^{2}$NTT, Inc., Japan \end{tabular}}

\begin{document}
\ninept
\maketitle

\begin{abstract}

We propose the Chunkwise Aligner, a novel architecture for streaming automatic speech recognition (ASR). 
While the Transducer is the standard model for streaming ASR, its training is costly due to the need to compute all possible audio-label alignments. 
The recently introduced Aligner reduces this cost by discarding explicit alignments, but this modification makes it unsuitable for streaming. 
Our approach overcomes this limitation by dividing the audio into chunks and aligning each label to the leftmost frames of its chunk, whereas transitions between chunks are managed by a learned end-of-chunk probability.
Experiments show that the Chunkwise Aligner not only matches the Transducer's accuracy in both offline and streaming scenarios, but also offers superior training and decoding efficiencies.

\end{abstract}
\section{Introduction}
State-of-the-art automatic speech recognition (ASR) systems are largely built on two dominant architectures: the Attention-based Encoder-Decoder (AED)~\cite{chorowski2015attention,vaswani2017attention} and the Transducer~\cite{graves2012sequencetransductionrecurrentneural}.
These architectures each offer distinct advantages. 
In this paper, 
we examine and compare them in terms of decoding speed, recognition accuracy, and streaming compatibility.

The AED is a powerful model trained in a label-synchronous manner with a simple cross-entropy objective, 
which makes its training process conceptually straightforward and efficient.
However, this simplicity comes at the cost of being inherently non-streaming, since its
decoder requires cross-attention over the entire audio sequence to predict each label. 
In contrast, and more relevant to our goal of streaming ASR, the Transducer is a frame-synchronous model
that naturally supports real-time applications~\cite{Sainath2021AnES}. 
This capability, however, relies on a more complex objective function, which requires a computationally expensive dynamic programming step to marginalize over all possible alignments between input audio frames and output labels.

A recent model, the Aligner~\cite{stooke2024aligner}, seeks to combine the strengths of both the Transducer and the AED. 
It adopts the Transducer architecture but removes the blank transition probability, and introduces a process called ``self-transduction'', 
in which the self-attention modules in the encoder learn to sequentially align all output labels to the leftmost frames of the input sequence.
Through this mechanism, the Aligner enables AED-like, label-synchronous training with a simple cross-entropy loss, whereas its decoding utilizes a Transducer-like, fast and lightweight autoregressive component.
However, by collapsing information to the beginning of the sequence, 
the model discards the local temporal alignment between the audio and the labels, which is essential for streaming recognition. 
Moreover, the Aligner is not robust to unseen speech lengths during inference and requires data concatenation to generalize across input lengths, thereby increasing its training cost.

\begin{figure}[t]
    \centering
    \includegraphics[width=\linewidth]{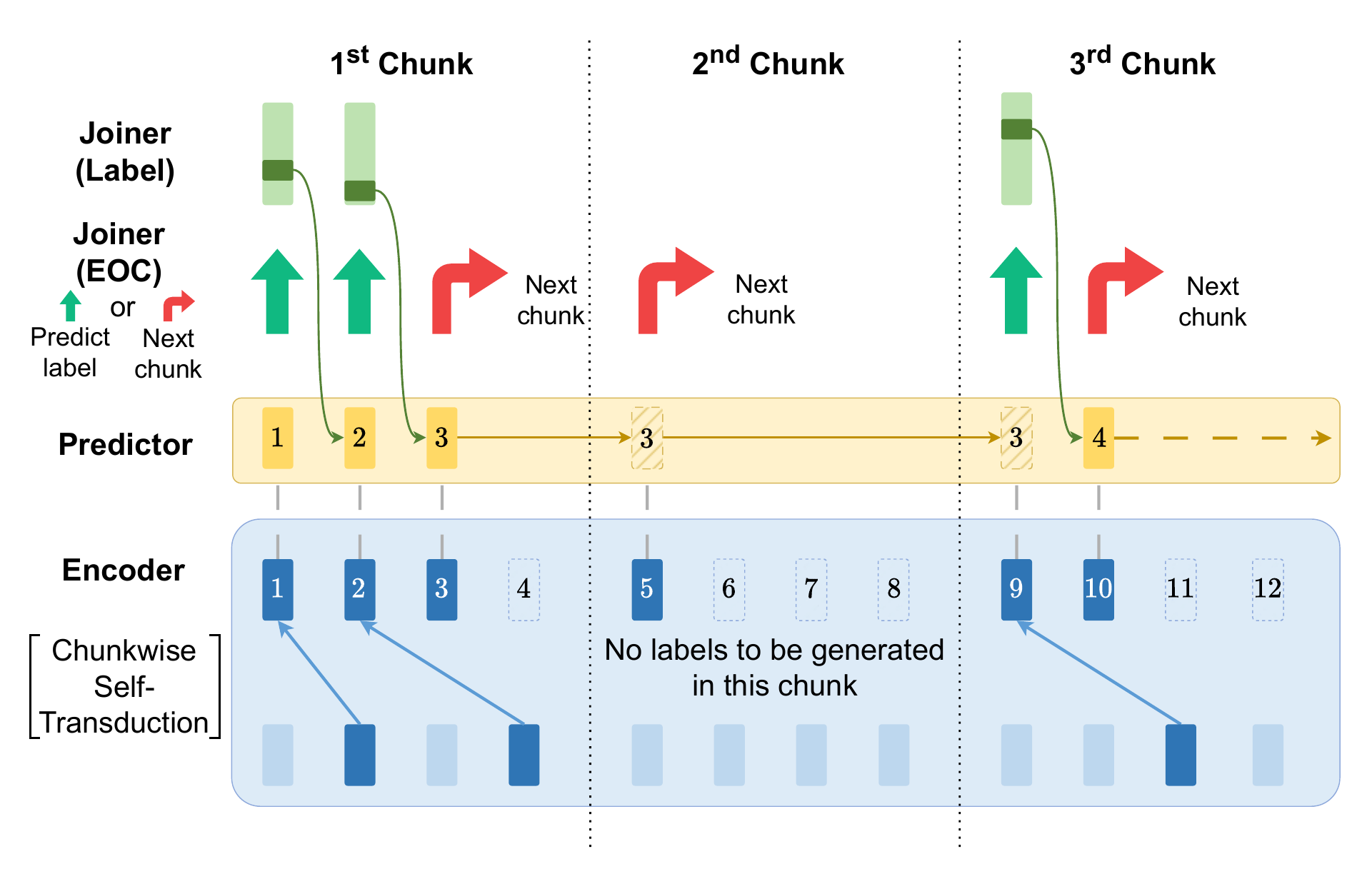}
    \vspace{-0.8cm}
    \caption{
        Schematic diagram of our proposed Chunkwise Aligner. 
        It predicts label probabilities and the EOC probability separately. 
        Chunkwise self-transduction performed by the encoder aligns label information (dense blue blocks) to the leftmost frames of each chunk.
        An EOC probability exceeding a predefined threshold advances the model into the next chunk,
        even if no labels were predicted (as is the case of the 2nd chunk).
    }
    \label{fig:concept}
\end{figure}

To address this limitation, we propose the \textit{Chunkwise Aligner}, 
a novel architecture that adapts the self-transduction mechanism for streaming ASR. 
As illustrated in \cref{fig:concept}, our model partitions the input audio into fixed-size chunks. 
Using a given alignment, it performs self-transduction locally by sequentially mapping the labels corresponding to each chunk to the leftmost frames within that chunk. 
To handle transitions between chunks, we introduce a learnable end-of-chunk (EOC) probability that indicates when to advance. 
Upon detecting an EOC, the hypotheses and decoder states of the current chunk are carried forward to the next, enabling the cycle to repeat for continuous audio streams.
Experimental results show that our Chunkwise Aligner not only surpasses the Aligner, but
achieves both offline and streaming ASR performances on par with that of the Transducer, 
while significantly reducing computational training costs and improving decoding speed.
\vspace{-0.2cm}
\section{Preliminaries} \label{sec:preliminary}
\vspace{-0.2cm}
The Transducer, Aligner, and our Chunkwise Aligner are all built upon a common encoder-predictor-joiner architecture. This section will first describe this common framework and then detail the joiner formulation, which differs between the Transducer and Aligner.

\vspace{-0.3cm}
\subsection{Model architecture}
\vspace{-0.1cm}
The architecture consists of three main components: an encoder, a predictor, and a joiner.
The processing of each component is described as follows:
\begin{itemize}
\item \textbf{Encoder} includes a self-attention module (e.g., Conformer~\cite{gulati2020conformer}) and transforms an input acoustic feature sequence $\bm{X} = \left[ \bm{x}_1, \dots, \bm{x}_{T} \right]$ of length $T$ into a $D^{\prime}$-dimensional higher-level representation $\bm{H}^{\text{enc}} = \left[ \bm{h}^{\text{enc}}_{1}, \dots, \bm{h}^{\text{enc}}_{T} \right] \in \mathbb{R}^{T \times D^{\prime}}$.

\item \textbf{Predictor} processes the previously emitted non-blank labels $Y = \left[ y_1, \dots, y_U \right]$ of length $U$, where $y_u \in \mathcal{Y}$, to autoregressively produce a sequence of $D^{\prime \prime}$-dimensional contextual embeddings $\bm{H}^{\text{pred}} = \left[ \bm{h}^{\text{pred}}_{1}, \dots, \bm{h}^{\text{pred}}_{U} \right] \in \mathbb{R}^{U \times D^{\prime \prime}}$. 
$\mathcal{Y}$ is the token set that consists of $V$ tokens. 

\item \textbf{Joiner} predicts the next label probabilities (additionally including a transition probability for the Transducer and Chunkwise Aligner) using $\bm{H}^{\text{enc}}$ and $\bm{H}^{\text{pred}}$.

\end{itemize}

\subsection{Joiner formulations}

\subsubsection{Transducer} \label{sssec:transducer}
The Transducer \cite{graves2012sequencetransductionrecurrentneural,variani2020hybrid} learns the mapping between sequences of different lengths through the introduction of blank symbol. 
Thus, the joiner computation must consider all possible alignments using the outputs of encoder and predictor. 
In this work, we employ the Hybrid Autoregressive Transducer (HAT) \cite{variani2020hybrid}, which produces the blank probability $\hat{y}_{t,u}^{\text{blank}} \in (0,1)^{1}$ and the label probabilities $\hat{\bm{y}}_{t,u}^{\text{label}} \in (0,1)^{V}$ separately. 
Here, we describe the details of the joiner processing in the HAT as follows:
\begin{eqnarray}
\bm{h}_{t,u}^{\text{joiner}} &=& \text{tanh} (\bm{W}^{\text{enc}} \bm{h}^{\text{enc}}_{t} + \bm{W}^{\text{pred}} \bm{h}^{\text{pred}}_{u} + \bm{b}), \label{eq:joiner_space_hat} \\
\hat{y}_{t,u}^{\text{blank}} &=& \text{Sigmoid} (\bm{w}^{\text{blank}} \bullet \bm{h}_{t,u}^{\text{joiner}} + b^{\text{blank}} ), \label{eq:blank} \\ 
\hat{\bm{y}}_{t,u}^{\text{label}} &=& \text{Softmax} (\bm{W}^{\text{label}} \bm{h}_{t,u}^{\text{joiner}} + \bm{b}^{\text{label}}), \label{eq:nonblank_hat} \\
\hat{\bm{y}}_{t,u} &=& \left[ \ \hat{y}_{t,u}^{\text{blank}}, \ (1 - \hat{y}_{t,u}^{\text{blank}}) \cdot \hat{\bm{y}}_{t,u}^{\text{label} \ \top} \ \right]^{\top} \in (0,1)^{V+1}, \label{eq:concat} 
\label{eq:hat}
\end{eqnarray}
where $\{ \bm{W}^{\text{enc}} \in \mathbb{R}^{D \times D^{\prime}}, \bm{W}^{\text{pred}} \in \mathbb{R}^{D \times D^{\prime \prime}}, \bm{b} \in \mathbb{R}^{D} \}$ is the parameter set of the joiner space, which produces $\bm{h}_{t,u}^{\text{joiner}}$ via the hyperbolic tangent function $\text{tanh}(\cdot)$. 
$\hat{y}_{t,u}^{\text{blank}}$ is computed using weight vector $\bm{w}^{\text{blank}} \in \mathbb{R}^{D}$ and bias term $b^{\text{blank}} \in \mathbb{R}^{1}$, where $\bullet$ and $\text{Sigmoid}(\cdot)$ denote the dot-product and sigmoid operation, respectively. 
$\{ \bm{W}^{\text{label}} \in \mathbb{R}^{V \times D}, \bm{b}^{\text{label}} \in \mathbb{R}^{V} \}$ are the parameters of the linear output layer, which yields $\hat{\bm{y}}_{t,u}^{\text{label}}$ via the softmax function $\text{Softmax}(\cdot)$.
Consequently, during training, the joiner computes a memory-intensive three-dimensional grid of predictions with shape $T \times U \times (V+1)$.
The Transducer loss is defined as the full-sum loss and can be computed efficiently using the forward-backward algorithm~\cite{graves2012sequencetransductionrecurrentneural}.

\vspace{-0.1cm}
\subsubsection{Aligner}
The Aligner \cite{stooke2024aligner} simplifies the joiner computation by enforcing a one-to-one mapping. 
It performs ``self-transduction'' using the self-attention module within the encoder, without employing cross-attention as in the AED. The self-transduction internally reorders acoustic information so that the leftmost $U$ frames of $\bm{H}^\text{enc}$ are aligned with the $U$ labels. 
Therefore, the joiner of the Aligner can easily compute the output probabilities $\hat{\bm{y}}^{\text{label}}_{u} \in (0,1)^{V}$ as follows:
\begin{eqnarray}
\bm{h}_{u}^{\text{joiner}} &=& \text{tanh} (\bm{W}^{\text{enc}} \bm{h}^{\text{enc}}_{u} + \bm{W}^{\text{pred}} \bm{h}^{\text{pred}}_{u} + \bm{b}), \label{eq:joiner_space_aligner} \\
\hat{\bm{y}}^{\text{label}}_{u} &=& \text{Softmax} (\bm{W}^{\text{label}} \bm{h}_{u}^{\text{joiner}} + \bm{b}^{\text{label}}). \label{eq:nonblank_aligner}
\label{eq:aligner}
\end{eqnarray}

Self-transduction enforces a fixed alignment, so the blank symbol for transitions is not required, and the joiner may compute a distribution over only the $V$ text tokens. 
The dense Transducer grid is replaced with a two-dimensional $U \times V$ tensor,
which is optimized by a cross-entropy loss.
However, this full-context reordering discards the original temporal correspondence between the audio and labels, making the Aligner inherently non-streaming.

\newcommand{\labelgenpair}{\mathcal{S}^\text{label}}
\newcommand{\chunktransitionpair}{\mathcal{S}^\text{transition}}

\section{Proposed Chunkwise Aligner} \label{sec:methodology}

Our Chunkwise Aligner performs chunkwise decoding, and its processing is illustrated in Fig.~\ref{fig:concept}. 
The encoder output sequence $\bm{H}^{\text{enc}}$ is segmented into $N$ chunks of length $L_{c}$, denoted as $(\bm{H}^{\text{enc}}_{1}, \dots, \bm{H}^{\text{enc}}_{N})$, where $\bm{H}^{\text{enc}}_{n} = \left[ \bm{h}^{\text{enc}}_{(n-1) \times L_{c} + 1}, \dots, \bm{h}^{\text{enc}}_{n \times L_{c}} \right]^{\top}$ represents the $n$-th input chunk. 
This enables the Chunkwise Aligner to operate in streaming ASR.
The predictor output sequence $\bm{H}^{\text{pred}}$ is also divided into $N$ chunks $(\bm{H}^{\text{pred}}_{1}, \dots, \bm{H}^{\text{pred}}_{N})$ to match the encoder output chunks.
Labels are assigned to chunks via forced alignment (similar to~\cite{speechllm}), thus the number of labels in each chunk varies.
Hence, the sequence of predictor outputs belonging to the $n$-th chunk is defined as 
$\bm{H}^{\text{pred}}_n = \left[ \bm{h}^{\text{pred}}_{u_n} \right]^\top_{u_n \in (1, ..., U_n)}$
where $U_n$ denotes the total number of labels assigned to the $n$-th chunk, and $u_n$ denotes the assigned position of the label within its chunk.
Whether to transition to the next chunk is determined by the end-of-chunk (EOC) probability, instead of the blank probability in the HAT joiner as described in~\cref{sssec:transducer}. 
The joiner of the Chunkwise Aligner has the same number of parameters as the HAT joiner and is defined as follows:
\begin{eqnarray}
\bm{h}_{u_{n}}^{\text{joiner}} &=& \text{tanh} (\bm{W}^{\text{enc}} \bm{h}^{\text{enc}}_{(n-1) \times L_{c} + u_{n}} + \bm{W}^{\text{pred}} \bm{h}^{\text{pred}}_{u_{n}} + \bm{b}), \label{eq:joiner_space_chunkwise_aligner} \\
\hat{y}_{u_{n}}^{\text{eoc}} &=& \text{Sigmoid} (\bm{w}^{\text{eoc}} \bullet \bm{h}_{u_{n}}^{\text{joiner}} + b^{\text{eoc}} ), \label{eq:eoc} \\
\hat{\bm{y}}_{u_{n}}^{\text{label}} &=& \text{Softmax} (\bm{W}^{\text{label}} \bm{h}_{u_{n}}^{\text{joiner}} + \bm{b}^{\text{label}}), \label{eq:nonblank_chunkwise_aligner} 
\label{eq:chunkwise_aligner}
\end{eqnarray}
where $\hat{y}_{u_{n}}^{\text{eoc}} \in (0,1)^{1}$ and $\hat{\bm{y}}_{u_{n}}^{\text{label}} \in (0,1)^{V}$ are the $u_{n}$-th EOC probability and label probabilities in the $n$-th chunk, respectively.
Thus, the self-attention modules in the Chunkwise Aligner's encoder are trained to perform the self-transduction mechanism of the Aligner in a chunkwise manner, i.e. ``chunkwise self-transduction''.

\subsection{Training objectives} \label{ssec:training}
The tensor format of the label probabilities is the same as that of the Aligner, $U \times V$, and they are optimized using a standard cross-entropy loss, $\mathcal{L}_\text{label}$. 
However, as shown in Fig.~\ref{fig:concept}, the EOC probabilities require one additional entry for each chunk, resulting in a prediction grid of $(U + N) \times 1$, which corresponds to the label length $U$ plus the chunk size $N$. 
For each chunk $n$, the $U_n$-th predictor output is paired with the $(U_n+1)$-th encoder output to compute the final EOC probability of that chunk. 
To train the EOC branch in the Chunkwise Aligner's joiner, we prepare $U+N$ target labels such that only the final position in each chunk is set to 1, while all others are set to 0. 
The resulting EOC probabilities are then optimized with a binary cross-entropy loss, $\mathcal{L}_\text{eoc}$. 
The final total loss, $\mathcal{L}_\text{total}$, is then a simple sum of these two components:
\begin{equation}
    \mathcal{L}_{\text{total}} = \mathcal{L}_\text{label} + \mathcal{L}_\text{eoc}.
\end{equation}

The prediction grids sum to a total size of $U \times \left(V + 1 + \frac{N}{U}\right)$, providing huge computational improvements compared to the Transducer, $T \times U \times (V + 1)$. 
Therefore, the Chunkwise Aligner drastically reduces memory and computational load during training, while retaining the streaming capability that the original Aligner lacks.

\begin{algorithm}[t]
\caption{
    Beam search decoding for the Chunkwise Aligner
}
\begin{algorithmic}[1]
\Function{BeamSearch}{$\bm{H}^{\text{enc}}, L_{c}, \text{beam\_size}, \tau $}
    \State \textbf{Input}: Encoder output, Chunk size, Beam size, Threshold
    \State \textbf{Output}: Most likely hypotheses
    \State $B \gets \{ (\texttt{<sos>}, 1, \bm{s}_0) \}$; $N \gets \lceil |\bm{H^{\text{enc}}}| / L_{c} \rceil$; $F \gets \{\}$ \label{algoline:initialize}
    \For{$n = 1, ..., N$} \Comment{Chunkwise processing} \label{algoline:chunkwise_processing}
        \State $\bm{H}^{\text{enc}}_n \gets \text{ChunkGenerator} (\bm{H}^{\text{enc}}, L_{c}, n ) $; $C \gets \{\}$ \label{algoline:initialize_outerloop}
        \For{ $\bm{h}_{\mathrm{t}}^\text{enc} \in \bm{H}^\text{enc}_n$ } \label{algoline:inner_loop}
            \State $A \gets \{\}$
            \If{$B$ is empty}
                \State \textbf{break} \label{algoline:break_innerloop}
            \EndIf
            \For {$(\mathbf{y}, \delta_{u-1}, \bm{s}_{u-1}) \in B$} \label{algoline:interloop_beamsearch}
                \State $\bm{h}^\text{pred}_u , \bm{s}_u \gets \text{Predictor}(y_{u-1}, \bm{s}_{u-1})$ \label{algoline:predictor_update}
                \State $\bm{h}^\text{joiner}_{u} \gets \text{Joiner}(\bm{h}^\text{enc}_{\mathrm{t}}, \bm{h}^\text{pred}_u)$ \label{algoline:joiner_space} \Comment{Eq. (\ref{eq:joiner_space_chunkwise_aligner})}
                \State $\hat{y}_{u}^{\text{eoc}} \gets \text{JoinerEOC}(\bm{h}^\text{joiner}_{u})$ \label{algoline:pred_eoc} \Comment{Eq. (\ref{eq:eoc})}
                \If{ $\hat{y}_{u}^{\text{eoc}} > \tau$} 
                    \State $C \gets C \cup \{(\mathbf{y}, \delta_{u-1} * \hat{y}_{u}^{\text{eoc}}, \bm{s}_{u-1})\} $ \label{algoline:transition_if_3}
                    \State \textbf{continue}
                \EndIf
                \State $\hat{\bm{y}}_{u}^{\text{label}} \gets \text{JoinerLabel}(\bm{h}^\text{joiner}_{u}) $ \label{algoline:pred_label} \Comment{Eq. (\ref{eq:nonblank_chunkwise_aligner})}
                \For{ $k \in \mathcal{Y}$ } \label{algoline:combination}
                    \State $\delta_{u} \gets \delta_{u-1} * \hat{\bm{y}}_{u}^{\text{label}} [k] * (1-\hat{y}_{u}^{\text{eoc}})$ \label{algoline:ypred_update_score}
                    \If{ $k$ is \texttt{<eos>}} 
                        \State $F \gets F \cup \{(\mathbf{y}+k, \delta_{u},\bm{s}_{u})\}$
                    \Else    
                        \State $A \gets A \cup \{(\mathbf{y}+k, \delta_{u},\bm{s}_{u})\}$       
                    \EndIf
                \EndFor
            \EndFor
            \State $B \gets \text{PruneHyps}(A, \text{beam\_size})$ \label{algoline:next_frame}
        \EndFor
        \State $B \gets \text{PruneHyps}(B \cup C, \text{beam\_size})$ \label{algoline:next_chunk}
    \EndFor
    \State \Return $\text{Sorted}(F)$ \label{algoline:sorted_return}
\EndFunction
\end{algorithmic}
\label{algo:inference}
\end{algorithm}

\subsection{Decoding strategy} \label{ssec:decoding}
\vspace{-0.1cm}
\cref{algo:inference} shows the chunkwise decoding process for the Chunkwise Aligner.
We prepare tentative and final hypothesis sets, $B$ and $F$, respectively (Line \ref{algoline:initialize}). 
$B = \{ \mathbf{y}, \delta, \bm{s} \} $ is initialized with the start-of-sentence label $\texttt{<sos>}$, a score of $1$, and the predictor state $\bm{s}_{0}$. 
From Line \ref{algoline:chunkwise_processing}, chunkwise decoding begins: the encoder output sequence $\bm{H}^{\text{enc}}$ is segmented into $N$ chunks, and an empty set $C$ is initialized to store hypotheses that have predicted a chunk transition (Line \ref{algoline:initialize_outerloop}). 
The inner loop (Line \ref{algoline:inner_loop}) performs a framewise beam search within the current chunk, $\bm{H}^\text{enc}_{n}$, 
until the end of the current chunk is reached or the tentative hypothesis set $B$ becomes empty.
From Line \ref{algoline:interloop_beamsearch}, for each hypothesis in the tentative hypothesis set $B$, the predictor and joiner computations are performed (Lines \ref{algoline:predictor_update}, \ref{algoline:joiner_space}, \ref{algoline:pred_eoc}, and \ref{algoline:pred_label}), and the resulting hypotheses are appended to one of the hypothesis sets $C$, $F$, or $A$. 
Note that during inference, the ground truth transcript is unknown, and we cannot calculate which chunk contains the $u$-th label beforehand. 
Instead, if $\hat{y}_{u}^{\text{eoc}}$ exceeds a threshold of $\tau=0.5$~\footnote{
    Preliminary experiments showed that varying the threshold $\tau$ did not yield a significant impact on recognition performance.}, we terminate the search for this chunk, and append that hypothesis to $C$ (Line \ref{algoline:transition_if_3}).
When computing the label probabilities $\hat{\bm{y}}_{u}^{\text{label}}$ (Line \ref{algoline:combination}), if the label $k$ is the end-of-sentence label \texttt{<eos>}, the corresponding hypothesis is appended to $F$; otherwise, it is appended to $B$.
$\text{PrunedHyps}(\cdot, \cdot)$ in Lines \ref{algoline:next_frame} and \ref{algoline:next_chunk} sorts the input hypothesis set by score and returns the top hypotheses, up to the beam size. 
If the decoding step reaches the end of the final chunk $N$, the hypotheses $F$, sorted by score, are returned (Line \ref{algoline:sorted_return}).

Compared to the Transducer, our Chunkwise Aligner enables a simple, label-synchronous-like algorithm that ensures decoding steps proportional to the hypothesis label length $U$ instead of the acoustic frame length $T$. 
Our decoding algorithm can reduce the computation of label probabilities $\hat{\bm{y}}_{u}^{\text{label}}$, which are computationally expensive, by thresholding $\tau$ 
(Line \ref{algoline:transition_if_3}). 
If all hypotheses are stored in $C$ (Line \ref{algoline:break_innerloop}), it is no longer necessary to compute up to the last frame of the chunk, leading to a further substantial reduction in computational cost. 
Our streaming Chunkwise Aligner does not require cross-attention and achieves faster decoding than chunkwise AED-based models~\cite{MoChA,TriggerAttention,StreamingAttention,Chunked_AED}, which are complex in both training and inference. 
Moreover, this also enables Chunkwise Aligner decoding to be faster than Aligner decoding, because the latter, like AED, requires a maximum expected length of the hypothesis, leading to more redundant computation~\cite{Hybrid_CTC_Attention,espnet}.

\vspace{-0.2cm}
\section{Experimental Evaluations} \label{sec:experiments}
\vspace{-0.2cm}

\subsection{Data} \label{ssec:data}
\vspace{-0.1cm}
We evaluated our proposed Chunkwise Aligner models on LibriSpeech~\cite{librispeech} and Corpus of Spontaneous Japanese (CSJ)~\cite{csj} using ESPnet~\cite{espnet}.
The input feature was an 80-dimensional log Mel-filterbank extracted with a 25ms window and a 10ms stride.
Augmentation methods~\cite{ko2015speedperturbation,specaugment,park2020adaptivespecaug} were applied during training. 
We adopted a word-piece tokenizer~\cite{sennrich2016bpe} with a vocabulary size of 1,000 for LibriSpeech, and a character-based tokenizer resulting in a vocabulary size of 3,262 for CSJ.
Ground truth alignments for Chunkwise Aligner training were generated using the Montreal Forced Aligner (MFA)~\cite{MFA}. 

\vspace{-0.2cm}
\subsection{System Configuration} \label{ssec:system}
\vspace{-0.1cm}
All models were built on the same encoder architecture.
We used 17 Conformer blocks with the same setup as Conformer-L~\cite{gulati2020conformer}, comprising approximately 110M parameters. 
An embedder of a two-layer 2-dimensional convolutional neural network (CNN) with 256 filters and a frame reduction rate of 4 was used before the Conformer blocks. 
The kernel size of 31 and the batch normalization within the convolution module were replaced with a kernel size of 15 and layer normalization.
In streaming mode, both the current and history encoder chunk sizes were set to 15, which
when considering the subsampling nature of our embedder, results in an algorithmic latency of 600ms. 
Depthwise convolution is replaced with a causal variant for the streaming encoder~\cite{chen2021lcconformer}.
The predictor consisted of a 640-dimensional LSTM~\cite{hochreiter1997lstm}. The decoding chunk size of Chunkwise Aligners was fixed at 15 in both offline and streaming modes. 
Additionally, for comparison, we built AED models with four Transformer decoder blocks, following~\cite{stooke2024aligner}.

The parameters of all models were initialized with random values.
But, as in~\cite{speechllm}, only the encoder was initialized with parameters pretrained with Connectionist Temporal Classification (CTC)~\cite{graves2006connectionist} when using CTC alignments for Chunkwise Aligner training.
This pretrained encoder also included an Inter-CTC loss~\cite{interctc} at the 10th layer, from which the alignments were generated. 
We used the Adam optimizer~\cite{Kingma2014adam} with a warm-up phase of 25k steps and a peak learning rate of 1.5e-3. Training was conducted for a total of 100 epochs. 
A beam size of 8 during inference was used in all models. 
We evaluated performance in terms of word error rate (WER) for LibriSpeech and character error rate (CER) for CSJ.
We also measured the real-time factor (RTF: \textit{decoding time} / \textit{data time}) for each offline model, tested on an Intel Xeon Gold 6430  3.4GHz CPU.

\vspace{-0.2cm}
\subsection{Results on LibriSpeech} \label{ssec:results_librispeech}
\vspace{-0.1cm}

\begin{table}[t]
    \centering
    \caption{
        WERs [\%] and RTFs 
        for \textbf{Offline ASR} on the LibriSpeech test sets. ``Alignment type'' shows the forced alignment method used.
    }
    \vspace{2pt}
    \label{tab:offline_results}
    \scalebox{0.95}[0.95]{
\begin{tabular}{lcccc}
\bhline{1pt} 
\multirow{2}{*}{}   & \multirow{2}{*}{\begin{tabular}[c]{@{}c@{}}Alignment \\ type\end{tabular}} & \multicolumn{2}{c}{WER}                               & \multirow{2}{*}{RTF} \\ \cline{3-4}
                    &                                                                            & \multicolumn{1}{l}{clean} & \multicolumn{1}{l}{other} &                      \\ \hline
Aligner (+DataConcat)~\cite{stooke2024aligner}  & -                                                                          & 2.3                       & 5.1                       & N/A                     \\ \hdashline
Transducer          & -                                                                          & 2.2                       & 4.9                       & 0.30                 \\
CTC                 & -                                                                          & 2.7                       & 6.7                       & 0.09                 \\
AED (+DataConcat)   & -                                                                          & 2.4                       & 5.4                       & 0.49                     \\
Aligner (+DataConcat) & -                                                                        & 2.4                       & 5.7                       & 0.18                     \\ \hline
\multirow{2}{*}{Chunkwise Aligner}   & ground-truth                                              & 2.2                       & 5.0                       & 0.12                 \\ 
                    & offline CTC                                                                & 2.2                       & 5.0                       & 0.12                 \\ \bhline{1pt} 
\end{tabular}
}
\end{table}

First, we present the offline ASR results on the LibriSpeech test sets in Table~\ref{tab:offline_results}, where 
the solid line separates the baselines (top) and our proposed models (bottom). 
The WERs for CTC were obtained from the pretrained model used for alignment generation. 
When training the AED and Aligner models, we randomly concatenated two utterances with a probability of 15\% (``+DataConcat''). 
This is to improve performance on the longer utterances in the test sets, as demonstrated in~\cite{stooke2024aligner}. 
Our reproduced Aligner model (+DataConcat) was not enough to achieve results reported in the literature~\cite{stooke2024aligner}.
We attribute this difficulty to the more challenging task of global self-transduction as the audio length increases,
requiring the encoder to memorize the positions of the leftmost frames and remap acoustic information through the self-attention space. 
Unlike the Aligner, our Chunkwise Aligner does not need data concatenation.

Comparing against the Transducer highlights two advantages of our method.
Despite its simple cross-entropy training, our Chunkwise Aligner achieved comparable results to the memory-intensive Transducer.
And, by virtue of our chunk transition mechanism with the EOC probability, it activates its predictor and joiner label-synchronously, 
leading to an inference time 2.6 times shorter than the Transducer.
For the offline scenario, we observe no performance difference between training with alignments generated from ground-truth and that from a pretrained CTC. 

\begin{figure}[t]
    \centering
    \begin{minipage}[b]{0.49\columnwidth}
    \centering
    \includegraphics[width=1.0\columnwidth]{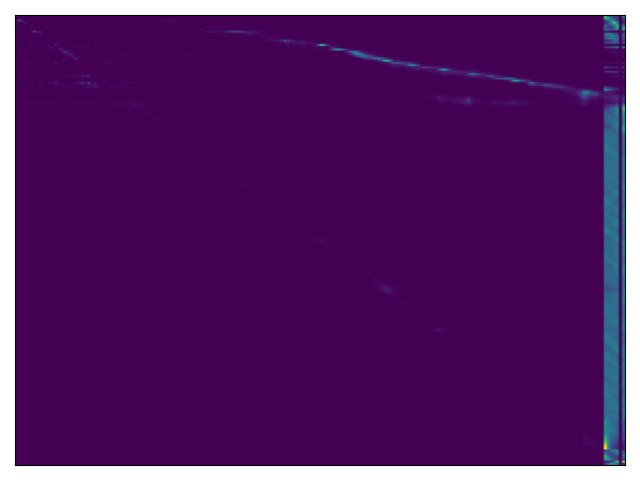}
    \\ (a) Aligner
    \label{fig:aligner}
\end{minipage}
\begin{minipage}[b]{0.49\columnwidth}
    \centering
    \includegraphics[width=1.0\columnwidth]{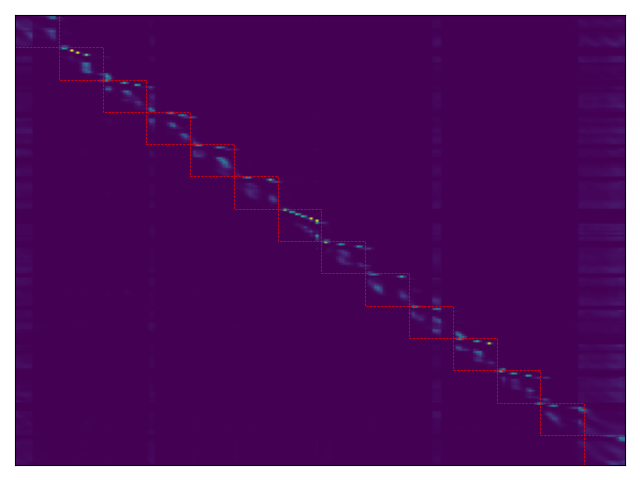}
    \\ (b) Chunkwise Aligner
    \label{fig:chunkwise_aligner}
\end{minipage}
    \vspace{-0.2cm}
    \caption{
        Self-attention weights from a single head at the 16th layer of (a) the Aligner and (b) the Chunkwise Aligner encoders performing audio-to-text alignment. 
        The red boxes in (b) mark the chunk boundaries along the diagonal.
        These figures are generated from the utterance ``1688-142285-0006'' (34 tokens).
    }
    \label{fig:alignment}
\end{figure}

We plotted the self-attention weights of the best offline Aligner and Chunkwise Aligner models in Fig.~\ref{fig:alignment}. 
Consistent with the findings in ~\cite{stooke2024aligner}, we find that the Aligner maps all labels sequentially to the leftmost frames of the utterance.
Meanwhile, our Chunkwise Aligner aligns labels on a chunkwise basis.
This shows that chunkwise self-transduction is a simpler objective because it maps the alignment over shorter distances, 
leading to robust generalization to utterances with unseen lengths and lower WERs.

\begin{table}[t]
    \centering
    \caption{
        WERs [\%] 
        for \textbf{Streaming ASR} on the LibriSpeech test sets. ``Delay'' is the delay duration applied to the forced alignments.
    }
    \vspace{2pt}
    \label{tab:streaming_results}
      \scalebox{0.95}[0.95]{
\begin{tabular}{lcccc}
\bhline{1pt} 
\multirow{2}{*}{} & \multirow{2}{*}{\begin{tabular}[c]{@{}c@{}}Alignment \\ type\end{tabular}} & \multirow{2}{*}{\begin{tabular}[c]{@{}c@{}}Delay\end{tabular}} & \multicolumn{2}{c}{WER}                               \\ \cline{4-5} 
                  &                                                                            &                                                                          & \multicolumn{1}{l}{clean} & \multicolumn{1}{l}{other} \\ \hline
Transducer        & -                                                                          & -                                                                        & 3.1                       & 7.6                       \\ 
CTC               & -                                                                          & -                                                                        & 4.1                       & 10.8                      \\ \hline
\multirow{6}{*}{\shortstack{Chunkwise \\ Aligner}} & ground-truth                                              & 0ms                                                                      & 3.9                       & 9.5                       \\
                  & ground-truth                                                               & 160ms                                                                    & 3.5                       & 8.5                       \\
                  & ground-truth                                                               & 320ms                                                                    & 3.2                       & 7.9                       \\
                  & ground-truth                                                               & 480ms                                                                    & 3.4                       & 8.3                       \\
                  & streaming CTC                                                              & 0ms                                                                      & 3.6                       & 9.0                       \\ 
\bhline{1pt} 
\end{tabular}
}
\vspace{-0.2cm}
\end{table}

Chunkwise self-transduction not only improves the performance of an offline Chunkwise Aligner, but unlike the Aligner, it also allows streaming ASR when a streaming encoder is adopted. 
Table~\ref{tab:streaming_results} presents our streaming ASR results. 
Compared to offline models, streaming ASR models commonly delay label emission. 
Therefore, the Chunkwise Aligner, which relies on forced alignments during training, may suffer from performance degradation due to mismatches in timing~\cite{lower_frame_rate,kd_streaming_ctc}. 
We investigated how the WERs are affected when the timestamps of ground truth forced alignments are delayed.
As can be seen, the best performance was achieved with a 320ms delay, yielding WERs comparable to the streaming Transducer.
Contrary to the offline scenario, we observed that using CTC alignments worsened performance. 
This is discussed in \cref{ssec:discussion}.

\vspace{-0.2cm}
\subsection{Results on CSJ} \label{ssec:results_other} 
\vspace{-0.1cm}

\begin{table}[t]
    \centering
    \caption{
        CERs [\%] and RTFs for the CSJ test sets.  
        Chunkwise Aligner models were trained using CTC alignments with no delay. 
    }
    \vspace{2pt}
    \label{tab:csj_results}
    \scalebox{0.95}[0.95]{
\begin{tabular}{lcccc|ccc}
\bhline{1pt} 
\multirow{2}{*}{} & \multicolumn{4}{c|}{Offline} & \multicolumn{3}{c}{Streaming} \\ \cline{2-5} \cline{6-8} 
                  & E1       & E2      & E3   &  RTF & E1       & E2       & E3      \\ \hline
Transducer        & 4.1      & 3.0     & 3.4 & 0.30 & 5.1      & 3.9      & 4.1     \\
CTC               & 4.2      & 3.1     & 3.6 & 0.10 & 5.3      & 4.2      & 4.4        \\
AED               & 3.9      & 2.9     & 3.4 & 0.55 & \multicolumn{3}{c}{N/A}       \\
Aligner           & 4.2      & 3.2     & 3.6 & 0.22 & \multicolumn{3}{c}{N/A}       \\ \hline
Chunkwise Aligner & 4.0      & 2.9     & 3.4 & 0.16 & 5.1      & 3.9      & 4.1        \\ \bhline{1pt} 
\end{tabular}
}
\end{table}

Shown in ~\cref{tab:csj_results}, our results for CSJ mirror those observed for LibriSpeech.
The Chunkwise Aligner again achieved CERs comparable to the Transducer in both offline and streaming settings,
while decoding 1.8 times faster in the offline scenario.
This finding demonstrate the cross-lingual applicability of our proposed method.

\vspace{-0.2cm}
\subsection{Discussion} \label{ssec:discussion} %
\vspace{-0.1cm}
Our results show that our method's performance is dependent on the alignment used during training.
We hypothesize that the degraded WERs on LibriSpeech when trained on CTC alignments is due to the subpar WERs of the streaming CTC model, from which the Chunkwise Aligner was initialized.
This hypothesis is supported by the strong results on CSJ by both the streaming CTC and Chunkwise Aligner models.
Improving performance without relying on high quality alignments remains a direction for future work.

\vspace{-0.2cm}
\section{Conclusion} \label{sec:conclusion}
\vspace{-0.2cm}
We proposed the Chunkwise Aligner, which improves the Aligner to support streaming through chunkwise processing. 
When compared to the Transducer, our method not only reduces training costs by relying entirely on cross-entropy training, but also decodes faster while maintaining comparable accuracy.
As future work, we plan to explore alignment-free training frameworks and investigate whether chunkwise aligners can work in attention-free encoders~\cite{moriya2025lcssm_dual,moriya2025all_in_one}.

\clearpage
\bibliographystyle{IEEEbib}
\bibliography{strings,refs}

\end{document}